\documentclass[%
twocolumn,
 amsmath,amssymb,
 aps,
pra,
]{revtex4-2}
\usepackage{orcidlink}
\usepackage{graphicx}
\usepackage{physics}
\usepackage{dcolumn}
\usepackage{bm}
\newcommand{\matr}[1]{\mathbf{#1}} 
\usepackage{blkarray}
\usepackage{amsmath,calc}

\begin{document}
\preprint{APS/123-QED}

\title{Complete description of fault tolerant quantum\\ gate operations for topological Majorana qubit systems}

\author{Adrian D. Scheppe\orcidlink{0000-0002-5566-2744}}\email{adrian.scheppe@afit.edu}
\author{Michael V. Pak}\email{michael.pak@afit.edu}
\affiliation{Department of Physics, Air Force Institute of Technology,\\
2950 Hobson Way, Wright-Patterson AFB, Ohio 45433}
\date{\today}

\begin{abstract}
Among the list of major threats to quantum computation, quantum decoherence poses one of the largest because it generates losses to the environment within a computational system which cannot be recovered via error correction methods. These methods require the assumption that the environmental interaction forces the qubit state into some linear combination of qubit eigenstates. In reality, the environment causes the qubit to enter into a mixed state where the original is no longer recoverable. A promising solution to this problem bases the computational states on the low lying energy excitations within topological materials. The existence of these states is protected by a global parameter within the Hamiltonian which prevents the computational states from coupling locally and decohering. In this paper, the qubit is based on non-local, topological Majorana fermions (MF), and the gate operations are generated by swapping or braiding the positions of said MF. The algorithmic calculation for such gate operations is well known, but, the opposite gates-to-braid calculation is currently underdeveloped. Additionally, because one may choose from a number of different possible qubit definitions, the resultant gate operations from calculation to calculation appear different. Here, the calculations for the two- and four-MF cases are recapitulated for the sake of logical flow. This set of gates serves as the foundation for the understanding and construction of the six-MF case. Using these, a full characterization of the system is made by completely generalizing the list of gates and transformations between possible qubit definitions. A complete description of this system is desirable and will hopefully serve future iterations of topological qubits.
\end{abstract}

\maketitle
\section{Introduction}
The current generation of superconducting (SC) qubits that form the basis of computation for state of the art systems has generated great strides for the field of quantum computation (QC). They satisfy the criteria for a ``good enough" qubit because one possesses a sufficient degree of control over the individual qubit state, and they scale better than other options \cite{scQubits, preskill1998quantum}.  However, there may exist an absolute ceiling for the scalability and tolerance of SC-qubit-based systems due to the problem of decoherence which particularly threatens the future of QC since the field deals with the manipulation of information \cite{ponnath2006difficulties}. One would hope to minimize these effects in order to maintain the integrity of the information provided to the computer. If the computer flipped a bit or deleted information without the user knowing, what good is this system?
\begin{figure}[b]
	\centering
	\includegraphics[width = \columnwidth]{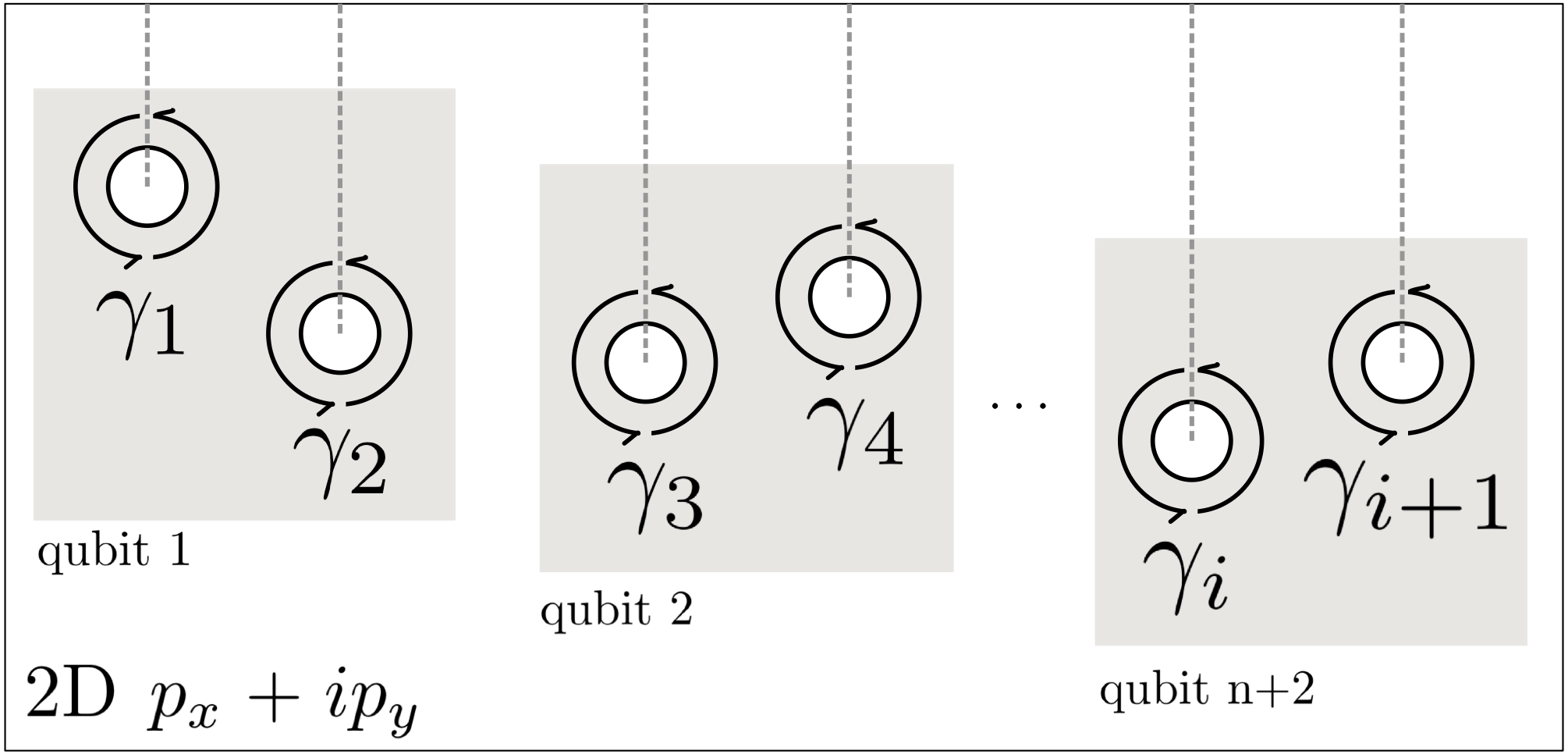}
	\caption[Majorana Fermion Setup]{Majorana fermion setup. The system is located in a 2D $p_x+ip_y$- wave SC which is induced on the surface of a TI by an \textit{s}-wave SC adjacent to the system. The qubits are defined by collecting two MF into one fermionic operator, depicted here as gray boxes. Vortices are made by allowing a magnetic field to penetrate the system with some strength in between the two critical values. Braids are made by switching positions of the vortices where one must cross the arbitrary branch cuts, depicted as dotted lines.}
	\label{fig:MFsetup}
\end{figure}

One may attempt to bolster the system by secluding the qubits from the environment as much as possible. Whatever decoherence that remains is dealt with by utilizing a number of error correction methods at one's disposal. However, these all require the assumption that the true state of the system remains a linear combination of qubit Hamiltonian energy eigenstates \cite{errorCorrect,errorCorrect2, errorCorrect3,errorCorrect4, errorCorrect5}. This would mean that the true state of the system is recoverable via a unitary transformation. In other words, we hope that the state remains pure when all is said and done. In actuality, environmental fluctuations force the qubit into a mixed state which is not at all a linear combination of the energy eigenstates of the qubit Hamiltonian \cite{ash2019study, decohere}.

For these reasons, an alternate solution to the decoherence problem has been proposed which makes use of topological states of matter \cite{kitaevOrigin}. As their name suggests, these are condensed matter systems equipped with a degenerate ground-state manifold based on some associated topology that is separated from the remaining spectrum by an energy gap \cite{wen2017colloquium,wenTop1}. If it is possible to construct a quantum computer based on a topologically invariant parameter, the computational states could not couple via local perturbations. Such a system is fault tolerant or decoherence proof \cite{nonabelianBigTwoColumn}.

The methods which outline direct calculation of unitary operators from a braid are well documented in several other publications \cite{nonabelianBigTwoColumn,sato2016majorana,lian2018topological,MF1,MF2,nonabelHalfVortex}. This paper supplements these works by providing the complete description of the computational space of the system. Such a description includes a completely generalized list of gate operations along with the transformations between possible qubit definitions. This description will assist the construction of the opposite directional algorithmic process, quantum gates to MF braids.

Here, the calculations for the two- and four-MF cases are recapitulated for the sake of logical flow. This set of gates will serve as the foundation for our understanding and construction of the gate operations for the six-MF case. Then, we will extend the list of gates to the general qubit and MF setting where we will list the general forms of all possible gates. Finally, we will demonstrate how one transforms the list of possible gates between different qubit definitions.
\section{Description of System}
There exist copious examples of well-studied systems with varying flavors and temperaments that exhibit topological qualities; however, it is not so common to discover one with the proper conditions for quantum computation.

A promising setting and the subject of this work makes use of nonlocal MF pairs within the induced 2D SC formed by adhering a 3D type-II, \textit{s}-wave SC to a strong topological insulator (TI) \cite{MFvortex} (see Fig. \ref{fig:MFsetup}).
The induced system is likewise type-II SC which is known to support local gap closures for points where the magnetic field is in between the first and second critical field values \cite{campbell1987superconductivity}. These may be thought of as pointwise boundaries which are accompanied by Abrikosov vortices, and, since the gap must close at these points, each vortex hosts a MF \cite{kitaevChain}.

These modes exist purely two dimensionally and, due to this fact, MF exchange statistics may be qualitatively different from 3D statistics. In three dimensions, $k$ indistinguishable particles have two choices of exchange: symmetrically (bosons) or antisymmetrically (fermions). However, 2D particles do not have the same constraints which force symmetric or antisymmetric exchange. 2D exchanges can in principle generate any phase in between $0$ and $\pi$, obeying anyonic statistics instead \cite{leinaas1977theory}. In some special non-Abelian cases these exchanges rotate a manifold spanned by the degenerate ground states. Exchanges within a system of $k$ non-Abelian anyons are described by the Braid group, $B_{k}$, which have unitary operator representations, $U^{(k)}_i$\cite{nonabelHalfVortex,nonabelianBigTwoColumn,MF2}.

This setting provides one of these special cases, and, as one vortex encircles another, the path taken generates a total Berry phase of $\pi$ to the state of the system. This winding number is represented by branch cuts depicted as dotted lines in Fig. \ref{fig:MFsetup}, starting at each vortex and ending somewhere on the borders of the region \cite{braidPfaffian}. These cuts are made arbitrarily and will not affect the total calculation as long as everything remains consistent.

One may only interact with these vortices through some macroscopic means \cite{conMFM,conOptical,conSTM,controlMech, ma2020braiding}, and the only actions one may take in regards to the MF operators, $\gamma_i$, are,
\begin{enumerate}
\item \textit{Relabelling: $\gamma_i\rightarrow\gamma_j$}
\item \textit{Crossing branch cuts: $\gamma_i\rightarrow-\gamma_j$}
\end{enumerate}
which are achieved by a physical exchange of MFs.
\section{Braid Calculation}
\subsection{Ground State Quasiparticle}
Working within the adiabatic limit, we assume that there are no quasiparticle excitations other than those at the Fermi level \cite{schrade2018majorana}. In other words, the only quasiparticles within our system are the MF. A pair of MF are simply a single ground state mode split into two locations in real space. In order for a single electron to occupy the ground state, it must magically split into two locations at once. 

The first step in gate calculation is to then redefine the ground state electron in terms of MF operators,
\begin{align*}
    a_n&=\frac{1}{\sqrt{2}}(\gamma_i+i\gamma_{i+1}),\\
    a_n^\dagger&=\frac{1}{\sqrt{2}}(\gamma_i-i\gamma_{i+1}),
\end{align*}
where each pair of $\gamma$'s is associated with $a$, a highly non-local fermionic operator \cite{nonabelHalfVortex}. We keep Fig. \ref{fig:MFsetup} in mind as we index each MF and fermion. We note here that the only requirement in making this definition is that each $\gamma$ is associated with one $a$ at a time. Aside from this, the definition that one makes is absolutely arbitrary. Transformations from one definition to another are discussed in the final section where it is shown that a definition transformation is simply a rotation of the coordinate system of the Bloch sphere representation of a single qubit.

The occupation of $a$, either $\ket{0}$ or $\ket{1}$, defines the computational space for this system. When one is ready to take a measurement of this system, bringing the vortices together removes the degeneracy in the ground state. The resulting two energy eigenstates are situated above and below the Fermi surface, and the occupation number associated with this operator represents the occupation of the final upper energy state once a fusion is made between the two MFs \cite{nonabelianBigTwoColumn}.

Since the thing that occupies that final state is a simple electron, the $a$ operators must obey regular fermion anticommutator rules, 
\begin{align}
\label{eq:antiComm1}
\{a_{n}, a^{\dagger}_{n}\}&=1,\\  
\{a_{n},a_{n}\}&=\{a^\dagger_{n},a^\dagger_{n}\}=0.
\label{eq:antiComm2}
\end{align}
From these relationships, the MF operators inherit their own anti commutation relationship \cite{nonabelHalfVortex},
\begin{equation}
\label{eq:antiComm}
 \{\gamma_{i},\gamma_{j}\}=2\delta_{i,j},
\end{equation}
yielding two rules for Majorana operators: (i) Two identical $\gamma$ operators in a row will annihilate to 1 and (ii) exchanging any two adjacent operators produces a negative sign.
\subsection{Braid group representation}
The representation that maps the braid group to linear operators, $\rho:B_{k}\rightarrow \mathcal{LO}$, is an exponentiation of the MF operators \cite{nonabelianBigTwoColumn,nonabelHalfVortex,braidPfaffian},
\begin{equation}
\label{eq:braidGen}
    \rho([b_i])=U_{i}=e^{\frac{\pi}{4}\gamma_{i}\gamma_{i+1}}=\frac{1}{\sqrt{2}}(1+\gamma_{i}\gamma_{i+1}).
\end{equation}
where a Taylor expansion of the exponential leads to the right-hand side of Eq. (\ref{eq:braidGen}). The specific use of $[b_i]$ instead of $b_i$ is meant to be precise with the fact that one cannot access any arbitrary braid from the braid group. We only have access to the equivalence classes of braids depicted in Fig. \ref{fig:braidGroup}. If the braid is deformable into one of these classes then it is not possible to determine the difference between them.

One may verify that this operator indeed has the correct action on the MF by performing a similarity transform on some arbitrary $\gamma_k$,
Since $\gamma_{i}\rightarrow\gamma_{i+1}$ and $\gamma_{i+1}\rightarrow -\gamma_{i}$, $U_i$ represents a counterclockwise braid where $\gamma_{i+1}$ crosses a branch cut as depicted in Fig. \ref{fig:braidGroup}.

\subsection{Two- and four-MF systems}
Defining $a_1$ in terms of $\gamma_{1,2}$ operators,
\begin{gather*}
 \begin{pmatrix} a_{1} \\ a^{\dagger}_{1} \end{pmatrix}
 =
  \frac{1}{\sqrt{2}}
  \begin{pmatrix}
   1 & i \\
   1 & -i
   \end{pmatrix}
   \begin{pmatrix}
   \gamma_{1}\\
   \gamma_{2}
   \end{pmatrix},
\end{gather*}
and, by taking the inverse,
\begin{gather*}
 \begin{pmatrix} \gamma_{1} \\ \gamma_{2} \end{pmatrix}
 =
  \begin{pmatrix}
   1 & 1 \\
   -i & i
   \end{pmatrix}
   \begin{pmatrix}
   a_{1}\\
   a^{\dagger}_{1}
   \end{pmatrix},
\end{gather*}
the MF operators may be expanded in terms of qubit operators. We then define the computational basis as $\{\ket{0}, \ket{1}\}=\{\ket{0},a^{\dagger}_{1}\ket{0} \}$. 
\begin{figure}[t]
	\centering
	\includegraphics[width =\columnwidth]{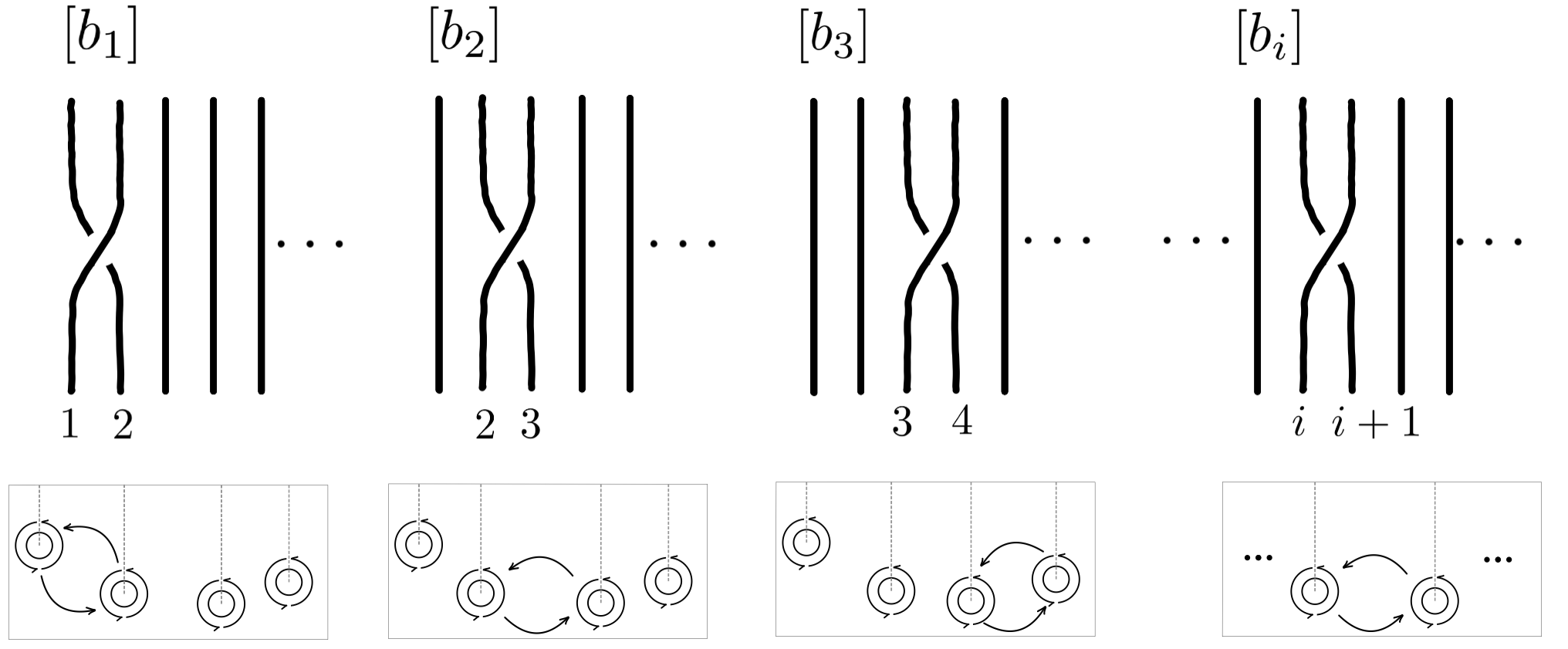}
	\caption[Braid group]{Braid Group. $B_k$ may be divided into equivalence classes, $[b_i]$, where the equivalence relation is defined as the braids which may be deformed into one another via some smooth operation. This system only has access to such classes of braids. Each braid is defined as the movement of one vortex around another.}
	\label{fig:braidGroup}
\end{figure}

To calculate the matrix representation of the only braid for a two MF system, $U^{(2)}_1$, simply rewrite Eq. (\ref{eq:braidGen}) in terms of the qubit operators, and operate on all members of the basis to determine the braid's effect on the ground state. In this document, the number within the  superscript parenthesis labels the number of MF's within the computational system, and the subscript labels the braid in accordance with Fig. \ref{fig:braidGroup}. The states then transform as,
\begin{align*}
    U^{(2)}_{1}\ket{0}&=\frac{1}{\sqrt{2}}(1+i)\ket{0},\\
    U^{(2)}_{1}\ket{1}&=\frac{1}{\sqrt{2}}(1-i)\ket{0}.
\end{align*}
In both cases, the operator is expanded in terms of fermionic operators, and the commutation rules, Eqs. (\ref{eq:antiComm1}) and (\ref{eq:antiComm2}) have been used  \cite{nonabelHalfVortex}. Dividing out a global phase factor of $e^{i\frac{\pi}{4}}$, $U^{(2)}_1$ in matrix form is,
\begin{gather*}
 U^{(2)}_{1}
 =\matr{S}=
  \begin{pmatrix}
   1 & 0 \\
   0 & i
   \end{pmatrix}.
\end{gather*}
This and the opposite braid, $U^{(2)\dagger}_1$ are the only two braids accessible to a two  MF system.

Using analogous methods from above, one may specify more gate operations by collecting more MF into the computational region of the 2D SC. For a four-MF system, form a new basis,
\begin{equation*}
    \{\ket{00}, \ket{01},\ket{10},\ket{11}\}=\{\ket{0},a^{\dagger}_{2}\ket{0},a^{\dagger}_{1}\ket{0},a^{\dagger}_{1}a^{\dagger}_{2}\ket{0} \},
\end{equation*}
and MF operators \cite{nonabelHalfVortex},
\begin{align*}
    \gamma_{1}&=a_1^{\dagger}+a_1,\\
    \gamma_{2}&=i(a_1^{\dagger}-a_1),\\
    \gamma_{3}&=a_2^{\dagger}+a_2,\\
    \gamma_{4}&=i(a_2^{\dagger}-a_2).
\end{align*}
Transforming the operators in an identical way to the two-MF case leads to linear operators,
\begin{gather*}
 U^{(4)}_{1}
 =
  \begin{pmatrix}
   1 & 0 & 0 & 0 \\
   0 & 1 & 0 & 0 \\
   0 & 0 & i & 0 \\
   0 & 0 & 0 & i \\
   \end{pmatrix}=\matr{S}\otimes \matr{I},
\end{gather*}
\begin{gather*}
 U^{(4)}_{2}
 =
 \frac{1}{\sqrt{2}}
  \begin{pmatrix}
   1 & 0 & 0 & i \\
   0 & 1 & i & 0 \\
   0 & i & 1 & 0 \\
   i & 0 & 0 & 1 \\
   \end{pmatrix}=\frac{1}{\sqrt{2}}(\matr{I}\otimes \matr{I}+i\matr{X}\otimes \matr{X}),
\end{gather*}
and,
\begin{gather*}
 U^{(4)}_{3}
 =
  \begin{pmatrix}
   1 & 0 & 0 & 0 \\
   0 & i & 0 & 0 \\
   0 & 0 & 1 & 0 \\
   0 & 0 & 0 & i \\
   \end{pmatrix}=\matr{I}\otimes \matr{S}.
\end{gather*}
One important observation is that the odd labeled braids, $U^{(4)}_1$ and $U^{(4)}_3$, involve MF defined under the same qubit operator while the only even labeled braid, $U^{(4)}_2$, shares MF from different qubits. This observation provides the qualitative difference between the equivalence classes of a given braid group. The specific scalar elements within the matrix are completely determined by how one defines the qubit operators from MF, but the ``shape" of each gate, i.e. diagonalized or coupling, are determined by whether or not MFs are shared between the qubits. This fact remains true regardless of the initial qubit definitions.

\subsection{Parity and subspaces}
The peculiar locations of non zero entries within $U^{(4)}_2$ are explained via parity, which, in this context, refers to the even or odd number of electrons in the SC bulk \cite{MF1,MF2,sarma2015majorana}.

Within a SC system, it is known that the electrons form Cooper pairs. If the system is sufficiently cooled and the gap is large enough, it is reasonable to expect that there must be an even number of electrons within the system. The only way that the system could potentially have an odd number of electrons would be if quasiparticle excitations were present. 

Since we base the computational space on counting particle numbers found within a given mode, this even-odd number rule constrains the computational space as well \cite{fu2010electron}. The number operator, $n = a_1^\dagger a_1=\frac{1}{2}(1+i\gamma_1\gamma_2)$, has two possibilities, 0 or 1, which also determines the even or odd nature of the total system. More specifically,
\begin{align*}
    \text{Even: }\frac{1}{2}(1+i\gamma_1\gamma_2)&=0
    \rightarrow i\gamma_1\gamma_2=-1,\\
    \text{Odd: }\frac{1}{2}(1+i\gamma_1\gamma_2)&=1
    \rightarrow i\gamma_1\gamma_2=1.
\end{align*}
These are the parity constraints placed upon the $\gamma$'s, and, since the braids are fermion conserving processes, the number of electrons remains unchanged throughout the braid. Therefore, the only states that are permitted to couple are those with equivalent parity \cite{MF1}. For this reason, a state with even parity, $\ket{11}$ cannot become a state of odd parity, $\ket{10}$. The $U^{(4)}_2$ transformation above demonstrates this property as even states become coupled with even states and odd states become coupled odd states. In other words, the final qubit in the ket, $\ket{00...\textbf{0}}$, couples with the parity state of the system \cite{sarma2015majorana}.

One then has a choice to work with states that include this last so called ancillary qubit or to exclude it. These two ways of performing calculations are called sparse or dense encoding, respectively \cite{sarma2015majorana,xu2011unified}. For a more rigorous approach and a general discussion of anyonic dense to sparse scheme transformations see Ref. \cite{xu2011unified}. It is typical to work in the dense scheme so that a more useful single qubit may be defined \cite{nonabelianBigTwoColumn,MF1}.  This is done by redefining the computational states as $\ket{\tilde{0}}\equiv\ket{00}$ and $\ket{\tilde{1}}\equiv\ket{11}$, and ignoring the odd subspace. The gates from above then become,
\begin{gather*}
 \tilde{U}^{(4)}_{1}= \tilde{U}^{(4)}_{3}
 =
   \begin{pmatrix}
  1 & 0 \\
  0 & i \\
\end{pmatrix}
=\matr{S},
\end{gather*}
and,
\begin{gather*}
 \tilde{U}^{(4)}_{2}
 =
 \frac{1}{\sqrt{2}}
   \begin{pmatrix}
  1 & i \\
  i & 1 \\
\end{pmatrix}
= \frac{1}{\sqrt{2}}(\matr{I}+i\matr{X}),
\end{gather*}
where the tilde denotes the even space reduced matrices. Once this convention is employed, a single qubit may be defined using four MF rather than two. 

These gates are decomposed above and shown to be intimately related to the phase gate and Bloch sphere \textit{x}-axis rotation operator,
\begin{equation*}
    R_{x}(\theta)=\cos(\frac{\theta}{2})\matr{I}-i\sin(\frac{\theta}{2})\matr{X},
\end{equation*}
evaluated for a specific rotational value, $\theta=-\frac{\pi}{2}$. When defining a qubit with four MF, this braid generates quarter counter clockwise rotations about the Bloch sphere's \textit{x}-axis, and, in combination with the phase gate, it is only possible for the qubit state to visit the six poles of the Bloch sphere. One may therefore create any of the quarter rotation Pauli $\matr{X}$, $\matr{Y}$, and $\matr{Z}$ gates. 
This single qubit case highlights that $\matr{S}$ and $\matr{R}_x(-\frac{\pi}{2})$ emerge naturally as native gates, a property which extends to higher dimensional computational spaces as well.
\subsection{Six MF system and beyond}
The two- and four-MF systems reveal a pattern regarding the calculation of even and odd braids. When braiding two MF from the same qubit, the resulting diagonal matrix imparts a phase factor on the $\ket{1}$ state defined within the qubit operator which contains the MF being swapped. For example, when braiding $\gamma_1$ and $\gamma_2$ defined under $a_1$, the computational states $\ket{1...}$ receive the phase factor $i$ and states $\ket{0...}$ are unaffected. The ellipses in the above kets are meant to demonstrate that it is irrelevant how the other states are populated when using the $U_1$ braid. Similarly, the coupling braids mix adjacent qubits in the ket. For example, the $U_2$ braid couples even states $\ket{00...}$ and $\ket{11...}$, or odd states $\ket{01...}$ and $\ket{10...}$ equivalently, with a phase factor of $i$ placed off diagonal. These observations allow one to easily construct the matrix form for any number of MF or qubits. 

Increasing the number of MF once more and implementing these observations, the compact forms of a six-MF system are,
\begin{align*}
    \tilde{U}^{(6)}_1&=\matr{S}\otimes \matr{I},\\
    \tilde{U}^{(6)}_{2}&=\frac{1}{\sqrt{2}}(\matr{I}\otimes\matr{I}+i\matr{X}\otimes\matr{X}),\\
    \tilde{U}^{(6)}_3&=\matr{I}\otimes \matr{S},\\
    \tilde{U}^{(6)}_{4}&=\frac{1}{\sqrt{2}}(\matr{I}\otimes\matr{I}+i\matr{I}\otimes\matr{X}),\\
    \tilde{U}^{(6)}_5&=\matr{S}\oplus i\matr{S}^{\dagger},
\end{align*}
where now patterns become noticeable for the compacted form of each gate as well. Aside from the final two ``ancillary" qubit braids, the odd braids contain a phase gate in the position of the tensor product that will only apply the gate to the qubit for which the braiding occurred. The even braids are similar in that the double tensor $\matr{X}$ product occurs in the tensor product in such a way that the $\matr{R}_x(\frac{\pi}{2})$ is applied only to the qubits which shared the MFs.

In other words, for a general $n$ qubit system, the nonancillary odd gates will fall into the following pattern,
\begin{align*}
    \tilde{U}^{(n+2)}_1&=\matr{S}\otimes \matr{I}\otimes \matr{I}\otimes\ldots,\\
    \tilde{U}^{(n+2)}_3&=\matr{I}\otimes \matr{S}\otimes \matr{I}\otimes\ldots,\\
    \tilde{U}^{(n+2)}_5&=\matr{I}\otimes \matr{I}\otimes \matr{S}\otimes\ldots,\\
    &\mathrel{\makebox[\widthof{=}]{\vdots}}
\end{align*}
and,
\begin{align*}
    \tilde{U}^{(n+2)}_2=\frac{1}{\sqrt{2}}(\matr{I}\otimes\matr{I}\otimes\ldots&+i\matr{X}\otimes\matr{X}\otimes\matr{I}\otimes\matr{I}\otimes\ldots),\\
    \tilde{U}^{(n+2)}_4=\frac{1}{\sqrt{2}}(\matr{I}\otimes\matr{I}\otimes\ldots&+i\matr{I}\otimes\matr{X}\otimes\matr{X}\otimes\matr{I}\otimes\ldots),\\
    \tilde{U}^{(n+2)}_6=\frac{1}{\sqrt{2}}(\matr{I}\otimes\matr{I}\otimes\ldots&+i\matr{I}\otimes\matr{I}\otimes\matr{X}\otimes\matr{X}\otimes\ldots),\\
    &\mathrel{\makebox[\widthof{-}]{\vdots}}
\end{align*}
The total number of gates within each tensor product is exactly equal to $n$, the number of non-ancillary qubits.


When using the sparse encoding, the ancillary braids will continue this pattern, but the dense encoding breaks out of the established pattern due to the fact that the even space reduction process partially excludes the states of the final qubit. The last even braid in a system, which shares MFs from the last data qubit and the ancillary, has the effect of applying the $\matr{X}$ gate to the final data qubit only,
\begin{align*}
    \tilde{U}^{(n+2)}_{final\ even}=\frac{1}{\sqrt{2}}(\matr{I}\otimes\ldots\otimes\matr{I}+i\matr{I}\otimes\ldots\otimes\matr{X}).
\end{align*}

The final odd braid, which swaps MF within the ancillary qubit, results in patterned tensor sum of the phase gate with its adjoint. This is easiest to demonstrate by lining the final odd gates side by side for the four, six, eight, and ten MF systems like so,
\begin{align*}
    \tilde{U}^{(4)}_3&=\matr{S},\\
    \tilde{U}^{(6)}_5&=\matr{S}\oplus i\matr{S}^{\dagger},\\
    \tilde{U}^{(8)}_7&=\matr{S}\oplus i\matr{S}^{\dagger}\oplus i\matr{S}^{\dagger}\oplus \matr{S},\\
    \tilde{U}^{(10)}_9&=\matr{S}\oplus i\matr{S}^{\dagger}\oplus i\matr{S}^{\dagger}\oplus \matr{S}\oplus i\matr{S}^{\dagger}\oplus \matr{S}\oplus \matr{S}\oplus i\matr{S}^{\dagger}.
\end{align*}
\section{Generalizing Gates}
To completely generalize these statements, for any $n\times n$ quantum gate acting on a system of $\frac{n}{2}$ qubits, one needs $n+2$ MFs for a dense encoding ($n$ for sparse). Including their undo action, there are then $2(n+1)$ types of equivalence class braids accessible. This collection of braids may be  subdivided into the $n+2$ odd, noncoupling gates and $n$ even, coupling gates. These braids only allow the multiqubit state to visit the six poles of their individual Bloch spheres. Therefore, unitary gates which place the single-qubit state anywhere in between these poles have no representation as a braid, and they are not possible with this implementation of qubit. Finally, one may condense the patterns for the $i$th braid observed into compact equations. For nonancillary braids ($i<n$), odd and even braids take the forms,
\begin{align*}
    \tilde{U}^{(n+2)}_i&=\matr{I}^{\otimes\frac{1}{2}(i-1)}\otimes \matr{S}\otimes \matr{I}^{\otimes\frac{1}{2}(n-i-1)},\\
    \tilde{U}^{(n+2)}_i&=\frac{1}{\sqrt{2}}(\matr{I}^{\otimes\frac{n}{2}}+i\matr{I}^{\otimes(\frac{i}{2}-1)}\otimes \matr{X}^{\otimes 2}\otimes \matr{I}^{\otimes\frac{1}{2}(n-i-2)}),
\end{align*}
respectively, and the ancillary ($i\geq n$) odd and even braids take the form,
\begin{align*}
    \tilde{U}^{(n+2)}_i&=\tilde{U}^{(n)}_{i-2}\oplus i\tilde{U}^{(n)\dagger}_{i-2},\\
    \tilde{U}^{(n+2)}_i&=\frac{1}{\sqrt{2}}(\matr{I}^{\otimes\frac{n}{2}}+i\matr{I}^{\otimes(\frac{i}{2}-1)}\otimes \matr{X}).
\end{align*}
These are compact statements of the previous section and represent the main findings of this work.

\section{Definition Transformation}
As mentioned previously, the qubits here are defined by pairing MF adjacently, but this choice is completely arbitrary. Using the sparse encoding for simplicity, let the collection of gates for a four-MF system with our definition be $D_0$, where,
\begin{equation*}
    D_0=\{ \matr{S}\otimes\matr{I},\matr{R}_{x}(-\frac{\pi}{2}),\matr{I}\otimes\matr{S}\},
\end{equation*}
and let an alternate definition, $D_1$, be to define $\gamma_1$, $\gamma_3$ and $\gamma_2$, $\gamma_4$ together. To change definitions, simply determine which braid from $D_0$ would move $\gamma_2$ into the $\gamma_3$ position and vice-versa. The braid which accomplishes this is $U_2$. Once this determination is made, perform a similarity transformation on the entire set $D_0$, $D_1=U^{\dagger}_2D_0U_2$,
\begin{gather*}
 U^{\dagger}_2U_1U_2
 =
 \frac{1}{\sqrt{2}}
  \begin{pmatrix}
   1 & 0 & 0 & -1 \\
   0 & 1 & 1 & 0 \\
   0 & -1 & 1 & 0 \\
   1 & 0 & 0 & 1 \\
   \end{pmatrix}=\frac{1}{\sqrt{2}}(\matr{I}\otimes \matr{I}-i\matr{X}\otimes \matr{Y}),
\end{gather*}
\begin{equation*}
 U_2^{\dagger}U_2U_2
 =U_2,
\end{equation*}
\begin{gather*}
 U^{\dagger}_2U_1U_2
 =
 \frac{1}{\sqrt{2}}
  \begin{pmatrix}
   1 & 0 & 0 & -1 \\
   0 & 1 & -1 & 0 \\
   0 & 1 & 1 & 0 \\
   1 & 0 & 0 & 1 \\
   \end{pmatrix}=\frac{1}{\sqrt{2}}(\matr{I}\otimes \matr{I}-i\matr{Y}\otimes \matr{X}).
\end{gather*}
This supports the general notion that qubit coupling will only occur when sharing MF between qubits. $D_1$ braids always couple since the neighboring MF are always from a different qubit. We also note that the dense encoding matrices reduce to $\matr{R}_{x}(-\frac{\pi}{2})$ and $\matr{R}_{y}(\frac{\pi}{2})$ now. These braids are the same as before; however, by changing the definition, we rotated the coordinate axis that the Bloch sphere is measured against. This makes the braids appear different.

Another alternate definition, $D_2$, could be where the middle and outer two MF are defined together. To acquire this definition from $D_0$, one would make the $D_0\rightarrow D_1$ transformation and then determine the braid from $D_1$ which swaps the new $\gamma_3$ and $\gamma_4$. In math, let $U_{i'}$ be the new gates under the $D_1$ definition,
\begin{align*}
    U_{3'}D_1U^{\dagger}_{3'}&=U_{3'}\{U_{2}D_0U^{\dagger}_{2}\}U^{\dagger}_{3'},\\
    &=U_{2}U_{3}\{U^{\dagger}_{2}U_{2}\}D_0\{U^{\dagger}_{2}U_{2}\}U^{\dagger}_{3}U^{\dagger}_{2},\\
    &=U_{2}U_{3}D_0U^{\dagger}_{3}U^{\dagger}_{2},
\end{align*}
where we have used unitarity of the operators in the last line. This transformation acts on the $D_0$ gates like so,
\begin{align*}
U_1&\rightarrow\frac{1}{\sqrt{2}}(\matr{I}\otimes \matr{I}-i\matr{X}\otimes \matr{Y}),\\
U_2&\rightarrow\matr{I}\otimes\matr{S}^\dagger,\\
U_3&\rightarrow\frac{1}{\sqrt{2}}(\matr{I}\otimes \matr{I}-i\matr{Y}\otimes \matr{X}),
\end{align*}
where the $U_2$ gate is the only one which does not share MF and is diagonal as expected.

It is even possible to switch the definitions of MF within each qubit and to swap qubit one and two using the same process, but the calculations are identical to the above math.
\bigskip
\section{Conclusion}
With a full description of braids and transformations between possible definitions, the computational space of this system is fully characterized. These results allow one to have better intuitions when implementing this example of a fault-tolerant system. Of course, the above gates do not form a universal set, but, as research into fault-tolerant systems continue, a better system may include either a full or partial implementation of MF. In order to use these qubits of the future, one will require a road map and intuitive understanding of the computational space.

It is more likely that the next generation of qubit designs will be some composite between conventional SC and topological qubits. This combination will hopefully lead to a more universal and robust qubit design. A complete description of this system is desirable and will hopefully serve future iterations of such qubits.
\bibliographystyle{apsrev4-2}
\bibliography{aps}
\end{document}